\documentclass{article}

\usepackage{arxiv}

\usepackage[utf8]{inputenc} % allow utf-8 input
\usepackage[T1]{fontenc}    % use 8-bit T1 fonts
\usepackage{hyperref}       % hyperlinks
\usepackage{url}            % simple URL typesetting
\usepackage{booktabs}       % professional-quality tables
\usepackage{amsfonts}       % blackboard math symbols
\usepackage{nicefrac}       % compact symbols for 1/2, etc.
\usepackage{microtype}      % microtypography
\usepackage{lipsum}		% Can be removed after putting your text content
\usepackage{graphicx}
\usepackage{multirow}
\usepackage{cite}

\usepackage{amsmath}
\usepackage{chemformula}

\usepackage{epstopdf}
\usepackage{array}
\usepackage{subfigure}
\usepackage{ragged2e}
\usepackage{booktabs,makecell,multirow,tabularx}
\usepackage[ruled,linesnumbered]{algorithm2e}

\title{Latent Residual-Closure Fourier Neural Operator for Robust Multi-Field Solving in Particle-in-Cell Simulations}

\author{Jianhua Lyu, ~Linlin Zhong\thanks{This paper is currently under consideration by a journal.} \\
	School of Electrical Engineering\\
	Southeast University\\
	No 2 Sipailou, Nanjing, Jiangsu Province 210096, China\\
	\texttt{linlin@seu.edu.cn}\\
}

\date{June 16, 2026}

\renewcommand{\headeright}{}
\renewcommand{\undertitle}{A Preprint}

\begin{document}
\maketitle

\begin{abstract}

	Particle-in-cell (PIC) simulations are widely used for kinetic plasma modeling in energy applications, but their efficiency is often limited by repeated field solves on dense meshes. This work proposes a Latent Residual-Closure Fourier Neural Operator (LRC-FNO) for robust surrogate multi-field solving in PIC simulations. Rather than treating field prediction as a purely data-driven regression task, LRC-FNO formulates PIC field solving as a two-level residual-closure problem involving source compression and source-to-field operator mapping. An autoencoder extracts compact representations of particle-deposited source fields, while a Latent Closure Refiner recovers unresolved residual structures lost during compression. A Coarse-FNO Solver captures the dominant field response, and a Residual-Closure FNO restores full-resolution corrections. The method is tested on three benchmarks with increasing complexity: 1D linear Landau damping (LLD), 2D two-stream instability (TSI), and a 2D scrape-off layer (SOL) fusion plasma model. In LLD and TSI, LRC-FNO better preserves charge-to-potential mapping, potential-mode evolution, residual charge structures, and particle-field energy exchange during closed-loop PIC integration. In the SOL case, LRC-FNO achieves relative $L_2$ errors of 0.0447 for the self-consistent potential and 0.0251 for the magnetic vector potential in single-step prediction. More importantly, when used as a neural initial guess with 20 iterative corrections, LRC-FNO maintains strong physical consistency in extrapolated closed-loop simulations, preserving charge and current density structures over a time range close to twice the training horizon. Direct network prediction and LRC-FNO-initialized iterative correction reduce the average single-step field-solution time from 499.65 ms to 14.99 ms and 28.17 ms, yielding field-solver speedups of 33.33× and 17.74×, respectively, while both strategies provide an overall PIC acceleration close to 2.5×. These results demonstrate that LRC-FNO can serve as both a fast surrogate field solver and a high-quality initialization strategy for iterative PIC field solvers.
\end{abstract}

\section{Introduction}
\label{sec:sec1}

\paragraph{}
The global energy transition has raised an urgent demand for predictive modeling and efficient simulation of complex energy-related systems. Plasma processes constitute an important class of strongly nonlinear and multiscale problems in this context, with representative applications including plasma-enabled energy conversion\cite{cite1,cite2}, eco-friendly gas insulation\cite{cite3,cite4,cite5}, and fusion plasmas\cite{cite6,cite7}. Particle-in-cell (PIC) simulation, as a first-principles kinetic method\cite{cite8}, has long been an important tool for plasma research because it can resolve sheath dynamics, wave-particle interactions\cite{cite9}, nonlinear instabilities\cite{cite10}, and self-consistent kinetic responses beyond fluid descriptions\cite{cite11}. However, the high computational cost of PIC simulations remains a major obstacle for long-time prediction, large-scale parametric studies, and digital-twin-oriented workflows\cite{cite12}.

\paragraph{}
In a standard PIC algorithm, particles are advanced under the Newton-Lorentz force, their charge and current are deposited onto the computational mesh, the field equations are solved on the mesh, and the resulting fields are interpolated back to particle positions8. The mesh-based field solve is a central component of the self-consistent particle-field coupling, since it converts deposited charge and current densities into the electromagnetic fields used in the next particle update. In electrostatic and magnetoquasistatic plasma models, the mesh-based field solve often involves the repeated solution of elliptic field equations, such as the mappings from charge density to electrostatic potential, $\rho$→$\phi$, and from current density to magnetic vector potential, $J$→$A$. As the simulation dimension, grid resolution, and boundary complexity increase, these repeated field solves can become a major computational bottleneck in the PIC cycle\cite{cite12}. Therefore, accelerating the field solver is a direct route to improving PIC efficiency. However, a surrogate field solver must preserve not only instantaneous field accuracy but also the stability of the coupled time integration. Otherwise, small field errors can propagate into subsequent particle motion and source deposition, interact with the intrinsic particle noise of PIC simulations, and become progressively amplified, eventually compromising the self-consistency of particle-field evolution and the long-time energy behavior.

\paragraph{}
This field-solve bottleneck has motivated the development of data-driven methods for accelerating partial differential equation (PDE) solution operators. In recent years, conventional machine learning methods\cite{cite13,cite14}, physics-informed neural network methods\cite{cite15,cite16}, and neural operator methods\cite{cite17,cite18} have been increasingly developed for accelerating PDE solutions. In particular, representative operator learning frameworks, such as DeepONet\cite{cite17}, FNO\cite{cite18}, GNO\cite{cite19}, and GINO\cite{cite20}, have enabled neural networks to learn mappings directly from input functions to output functions, and have gradually become important tools for accelerating field equation solvers in fluid dynamics, magnetohydrodynamics, and electromagnetic simulations. Existing neural Poisson solvers have also demonstrated the feasibility of learning field solutions on structured grids, but most of them focus on direct regression from smooth continuum source fields and do not address noisy particle-deposited source fields or closed-loop PIC stability\cite{cite21,cite22}. Therefore, directly applying existing neural operators to surrogate field solving in PIC simulations remains challenging. Unlike the smooth source fields commonly encountered in continuum models, the charge and current densities in PIC simulations are deposited from a finite number of particles and therefore naturally contain particle noise, local oscillations, and discontinuous features\cite{cite23,cite24,cite25}. Meanwhile, a single PIC case usually contains hundreds to thousands of time steps, and parametric studies can generate a large number of temporally correlated source-solution snapshots for each fixed set of physical parameters. When Courant-Friedrichs-Lewy (CFL) condition, Debye-length resolution requirements, or sheath-scale structures are involved, even finer temporal and spatial resolutions may be required\cite{cite26,cite27}. Therefore, a PIC surrogate field solver must not only learn a robust operator mapping from noisy source fields to solution fields, but also preserve field structures, boundary responses, and long-time coupling stability on high resolution grids. Directly feeding high-dimensional source fields into neural networks can make the model sensitive to particle noise and local high-frequency fluctuations, while also increasing the training and inference burden.

\paragraph{}
Reduced order models (ROMs) provide another important route for accelerating high-dimensional physical simulations. Classical methods such as proper orthogonal decomposition (POD) and dynamic mode decomposition (DMD) identify dominant coherent structures in physical fields and represent the original high-dimensional system in a low-dimensional subspace, thereby approximating system evolution at a much lower computational cost\cite{cite28}. Although continuum models and particle-based models differ fundamentally in their modeling paradigms and discrete representations, this low-dimensional representation of physical fields remains instructive for constructing surrogate field solvers in PIC simulations. For Vlasov-Poisson systems, existing studies have shown that POD-, DMD-, and autoencoder-based reduced representations can reduce computational cost while retaining the dominant dynamics, and can be further combined with neural networks or operator-learning methods to improve PDE surrogate solving\cite{cite29,cite30,cite31,cite32,cite33}. However, ROM-based solution operators, including our previous work\cite{cite34}, still face a central challenge, since the projection or encoding process inevitably truncates unresolved modes and local details, causing the reduced operator learned from low-dimensional representations to deviate from the true source-to-field mapping of the original high-dimensional system, especially under complex boundary conditions. In PIC simulations, this issue can be further amplified because source field representation and source-to-field operator learning are strongly affected by particle noise, local oscillations, and boundary responses. To compensate for truncation errors and unresolved dynamics in ROMs, closure modeling has become an important topic in ROMs research28. Recent studies have developed data-driven closure and correction strategies from different perspectives, including pressure data-driven VMS-ROMs\cite{cite35}, data-driven multiscale stochastic ROM frameworks\cite{cite36}, and symbolic-regression-based ROM closures for under-resolved convection-dominated flows\cite{cite37}.

\paragraph{}
Motivated by these developments, we propose a Latent Residual-Closure Fourier Neural Operator, termed LRC-FNO, for robust multi-field surrogate solving in PIC simulations. Rather than treating field prediction as a purely data-driven regression task, LRC-FNO formulates PIC field solving as a residual-closure problem, in which the residuals are not arbitrary neural corrections but unresolved components introduced by source field compression and coarse source-to-field operator approximation. The proposed framework introduces residual closure into the neural operator field solver at two levels. First, an autoencoder maps high-dimensional source fields into a ROM-like latent representation, and a Latent Closure Refiner constructs a latent-space closure to compensate for information loss during source field reduction. Second, a Coarse-FNO Solver predicts the dominant field response based on the closure-enhanced latent representation, while a Residual-Closure FNO learns a field-space residual closure to correct unresolved residuals in the coarse solution. Through this two-level closure design, LRC-FNO aims to improve the robustness of source field encoding, the accuracy of the source-to-field operator mapping, and the consistency of closed-loop particle-field time integration. Therefore, LRC-FNO is designed not only to reduce the cost of field solves, but also to better preserve long-time energy behavior when embedded into coupled particle-field simulations. The framework is applied to representative electrostatic and magnetoquasistatic field-solve problems in PIC, including mappings from charge density to electrostatic potential and from current density to magnetic vector potential, and is evaluated not only by field prediction errors but also by long-time PIC diagnostics.

\paragraph{}
Our main contributions are summarized as follows:
\begin{itemize}

\item A Latent Closure Refiner is introduced to enhance autoencoder-based source field representations, improving the representation of noisy and boundary-sensitive PIC source fields.

\item A Coarse-FNO Solver is constructed using closure-enhanced compressed source fields, which enables the neural operator to capture dominant physical structures while reducing the influence of high-dimensional particle noise in PIC-oriented field solving.

\item A Residual-Closure FNO is developed to correct unresolved field components after the coarse-solve, compensating for fine-scale and high-frequency residuals on the full-resolution grid, enabling high-fidelity approximation of the source-to-field solution operator.
\end{itemize}
\paragraph{}
The rest of the paper is organized as follows. Section 2 introduces the PIC simulation framework and the source-to-field mappings considered in this work. Section 3 presents the proposed LRC-FNO architecture. Section 4 reports three numerical benchmarks, including 1D linear Landau damping, 2D two-stream instability, and a 2D scrape-off layer model. Finally, conclusions are drawn in Section 5.

\section{Particle-in-cell Simulation}
\label{sec:sec2}
\paragraph{}
Particle-in-cell (PIC) simulation describes the kinetic evolution of charged particles coupled with self-consistent electromagnetic fields. Instead of treating the plasma as a continuous fluid, PIC represents the phase-space distribution by a finite number of macroparticles, typically ranging from tens of thousands to hundreds of thousands or more depending on the simulation scale. In a standard PIC cycle, these computational particles are advanced in phase space, their charge and current are deposited onto a mesh, the field equations are solved on the mesh, and the resulting fields are interpolated back to the particle positions for the next update, as shown in Figure 1. In this work, we focus on the mesh-based field-solve component, which maps particle-deposited source fields to the corresponding electromagnetic potentials.

\begin{figure}[!b]
	\centering
	\includegraphics{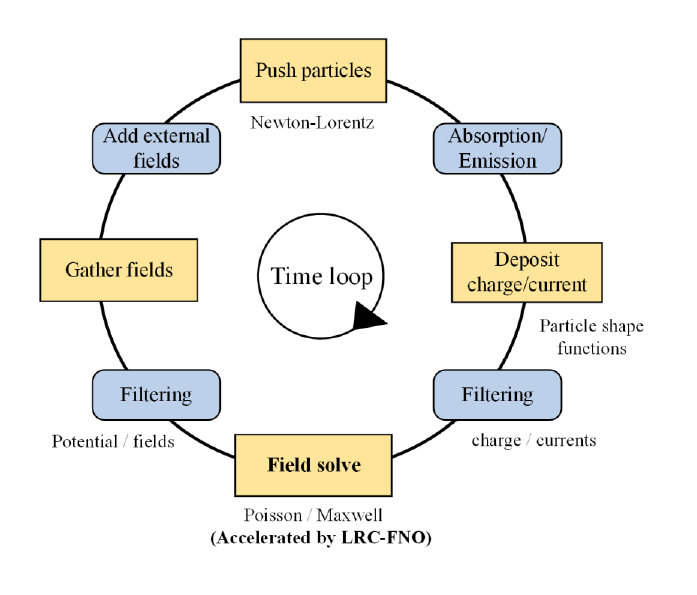}
	\caption{Schematic of the particle-in-cell cycle.}
	\label{fig:fig1}
\end{figure}

\paragraph{}
For a plasma consisting of multiple species, the charge density and current density deposited on the computational mesh are given by

\begin{equation}
	\label{equ:equ1}
	\rho \left( {\bf{x}} \right) = \sum\limits_s {\sum\limits_{p \in s} {{q_s}S\left( {{\bf{x}} - {{\bf{x}}_p}} \right)} } 
\end{equation}
\begin{equation}
	\label{equ:equ2}
    {\bf{J}}\left( {\bf{x}} \right) = \sum\limits_s {\sum\limits_{p \in s} {{q_s}{{\bf{v}}_p}S\left( {{\bf{x}} - {{\bf{x}}_p}} \right)} } 
\end{equation}

where $s$ denotes the particle species, $p$ indexes the computational particles belonging to species $s$, $q_s$ is the particle charge, and $\mathbf{x}_p$ and $\mathbf{v}_p$ are the particle position and velocity, respectively. $S$ is the particle shape function used for charge and current deposition. In this work, a first-order linear particle shape is used. In one dimension, it is written as

\begin{equation}
	\label{equ:equ3}
    S\left( x \right) = \left\{ {\begin{array}{*{20}{c}}
		{1 - \frac{{\left| x \right|}}{{\Delta x}},\left| x \right| < \Delta x}\\
		{0,\left| x \right| \ge \Delta x}
\end{array}} \right.
\end{equation}

\paragraph{}
For multidimensional simulations, the particle shape function is constructed by the tensor product of the corresponding one-dimensional shape functions. Through this deposition procedure, the discrete particle distribution is converted into mesh-based source fields, such as $\rho$ and $\mathbf{J}$.

\paragraph{}
For electrostatic field solves, the scalar potential $\phi$ is obtained from Poisson’s equation:

\begin{equation}
	\label{equ:equ4}
{\nabla ^2}\phi  =  - \frac{\rho }{{{\varepsilon _0}}}
\end{equation}
where $\varepsilon_0$ is the vacuum permittivity. For magnetoquasistatic field solves, the magnetic vector potential $\mathbf{A}$ is obtained from

\begin{equation}
	\label{equ:equ5}
{\nabla ^2}{\bf{A}} =  - {\mu _0}{\bf{J}}
\end{equation}
where $\mu_0$is the vacuum permeability. The electric and magnetic fields are then computed from the scalar and vector potentials as

\begin{equation}
	\label{equ:equ6}
{\bf{E}} =  - \nabla \phi 
\end{equation}
\begin{equation}
	\label{equ:equ7}
	{\bf{B}} = {{\bf{B}}_{{\rm{ext}}}} + \nabla  \times {\bf{A}}
\end{equation}

where $\mathbf{B}_{ext}$ denotes a prescribed external magnetic field when present.

\paragraph{}
Particle motion is advanced according to the Newton-Lorentz equations:

\begin{equation}
	\label{equ:equ8}
\frac{{d{{\bf{x}}_p}}}{{dt}} = {{\bf{v}}_p}
\end{equation}
\begin{equation}
	\label{equ:equ9}
	{m_s}\frac{{d{{\bf{v}}_p}}}{{dt}} = {q_s}\left[ {{\bf{E}}\left( {{{\bf{x}}_p}} \right) + {{\bf{v}}_p} \times {\bf{B}}\left( {{{\bf{x}}_p}} \right)} \right]
\end{equation}

where $m_s$ is the particle mass. The mesh-based fields are interpolated to particle positions using the same particle shape function as that used for deposition. In the numerical implementation, particle motion is advanced using a standard Boris scheme\cite{cite8}, which is widely used in PIC simulations because of its robustness for charged-particle motion in electromagnetic fields.

\paragraph{}
Equations (4) and (5) define the two general source-to-field mappings considered in this work. These mappings constitute the field-solve targets of the proposed neural surrogate. The electrostatic mapping $\rho$→$\phi$ is used in the  electrostatic benchmark cases, while the vector-potential mapping $\mathbf{J}$→$\mathbf{A}$ is considered in the magnetoquasistatic field-solve problem. The specific nondimensional forms and component-wise implementations used in each numerical benchmark are described in the corresponding case sections.

\section{Latent Residual-Closure Fourier Neural Operator}
\label{sec:sec3}
\paragraph{}
Building on the analysis of PIC source field characteristics and the need for surrogate field solving, this section presents the Latent Residual-Closure Fourier Neural Operator (LRC-FNO) for efficient field prediction in PIC simulations. The objective of LRC-FNO is to learn a robust operator mapping from particle-deposited source fields to the corresponding solution fields without directly using the raw particle-deposited full-resolution source fields as neural-operator inputs. This design reduces the influence of high-dimensional source representations and closed-loop particle-field coupling on model stability. The overall architecture of LRC-FNO is shown in Figure 2. The framework introduces residual closure at two coupled levels. 

\begin{figure}[!b]
	\centering
	\includegraphics[width=16cm]{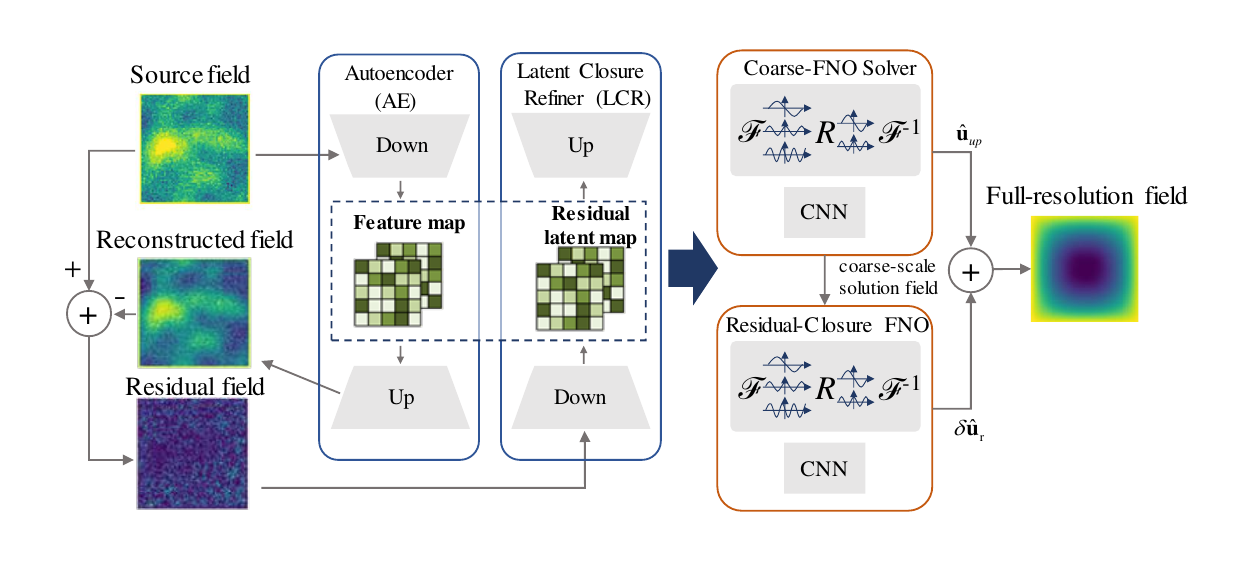}
	\caption{Overall architecture of LRC-FNO with latent-space and field-space residual closures.}
	\label{fig:fig2}
\end{figure}

\paragraph{}
The first level is the latent-space residual closure for source field representation. An autoencoder first extracts a ROM-like low-dimensional latent representation of the particle-deposited source field, while the Latent Closure Refiner (LCR) further learns the residual information that is not fully represented by the autoencoder. This produces a closure-enhanced latent source representation containing both dominant low-dimensional structures and unresolved local residual features.

\paragraph{}
The second level is the field-space residual closure for source-to-field operator learning. The closure-enhanced latent representation is first decoded into a task-oriented source proxy, which is then passed to a Coarse-FNO Solver to predict the dominant coarse-scale field response. The coarse solution is subsequently upsampled to the original grid, and a Residual-Closure FNO learns the full-resolution correction. In this way, the proposed method decomposes field prediction into coarse-scale operator learning and full-resolution residual closure, rather than directly predicting the complete high-resolution solution field in a single step.

\subsection{Latent-space residual closure with autoencoder and Latent Closure Refiner}
\label{sec:sec3.1}
\paragraph{}
The first closure level of LRC-FNO is the latent-space residual closure, which aims to construct a compact and information-enriched representation of particle-deposited source fields. To handle fine-scale source field structures dominated by particle noise, an autoencoder is first used to extract a ROM-like latent representation, while a Latent Closure Refiner is further introduced to compensate for the local residual information truncated during AE compression.

\paragraph{}
Let $\mathbf{\tilde{s}} \in \mathbb{R}^{C_s \times H \times W}$ denote a normalized source field sample, where $C_s$ is the number of source channels, and $H$ and $W$ are the spatial grid sizes.The encoder $\mathcal{E}_{\theta}$ maps $\tilde{\mathbf{s}}$ into a low-dimensional AE feature map:

\begin{equation}
	\label{equ:equ10}
	{{\bf{z}}_{{\rm{AE}}}} = {{\cal E}_\theta }({\bf{\tilde s}}),\qquad {{\bf{z}}_{{\rm{AE}}}} \in {\mathbb{R}^{{C_{{\rm{AE}}}} \times {H_l} \times {W_l}}}
\end{equation}

where $C_AE$ is the channel number of the AE feature map, and $H_l$ and $W_l$ are the spatial sizes of the latent grid. 

\paragraph{}
The decoder $\mathcal{D}_{\theta}$ reconstructs the source field from this latent representation.

\begin{equation}
	\label{equ:equ11}
  	{{\bf{\hat s}}_{{\rm{AE}}}} = {{\cal D}_\theta }\left( {{{\bf{z}}_{{\rm{AE}}}}} \right),\qquad {{\bf{\hat s}}_{{\rm{AE}}}} \in {\mathbb{R}^{{C_s} \times H \times W}}
\end{equation}
\paragraph{}
The autoencoder acts as a nonlinear reduced-order representation and captures the dominant low-dimensional structure of the source field. In the first training stage, the autoencoder is optimized using the reconstruction loss:

\begin{equation}
	\label{equ:equ12}
	{{\cal L}_{{\rm{AE}}}} = \parallel {{\bf{\hat s}}_{{\rm{AE}}}} - {\bf{\tilde s}}\parallel _2^2
\end{equation}

\paragraph{}
However, the compression process inevitably discards part of the unresolved local physical information. For PIC source fields, these truncated components may still contain physically relevant density or current fluctuations near boundaries and strong-gradient regions. Although filtering and reduced representations are commonly used to suppress particle noise in charge and current fields, an overly smoothed source representation may introduce nonphysical errors when embedded into closed-loop PIC time integration. Therefore, a Latent Closure Refiner is introduced to recover the residual information lost during AE compression. Instead of directly re-encoding the original source field, the Latent Closure Refiner operates on the AE reconstruction residual field:

\begin{equation}
	\label{equ:equ13}
	{{\bf{r}}_{{\rm{AE}}}} = {\bf{\tilde s}} - {{\bf{\hat s}}_{{\rm{AE}}}},\qquad {{\bf{r}}_{{\rm{AE}}}} \in {\mathbb{R}^{{C_s} \times H \times W}}
\end{equation}

\paragraph{}
This residual represents the locally unresolved component that is not fully captured by the AE latent space. Therefore, the Latent Closure Refiner can be interpreted as a residual closure model in the source field latent space.

\paragraph{}
As shown in Figure 3, the Latent Closure Refiner adopts a patch-wise latent representation. The residual field $\mathbf{r}_{AE}$ is first partitioned into local patches and projected into a latent feature space by a convolutional patch embedding operator:

\begin{equation}
	\label{equ:equ14}
    {{\bf{h}}_0} = {\cal P}\left( {{{\bf{r}}_{{\rm{AE}}}}} \right),\qquad {{\bf{h}}_0} \in {\mathbb{R}^{{C_e} \times {H_l} \times {W_l}}}
\end{equation}

where $\mathcal{P}$ denotes the convolutional patch embedding operator, and $C_e$ is the embedding channel number. This operation embeds local residual regions into patch-wise latent elements and forms a two-dimensional residual latent map corresponding to the AE latent grid. The design is inspired by patch embedding and positional embedding in vision models, such as Vision Transformer\cite{cite38}. Unlike Transformer token representations based on self-attention, the embedded patches here are treated as local representation elements in the latent space. Inspired by attention-free patch-wise representations such as MLP-Mixer\cite{cite39}, a Latent MLP encoder is used to perform nonlinear transformation on these patch-wise latent elements and obtain a compact residual latent representation.

\begin{figure}[!b]
	\centering
	\includegraphics[width=15cm]{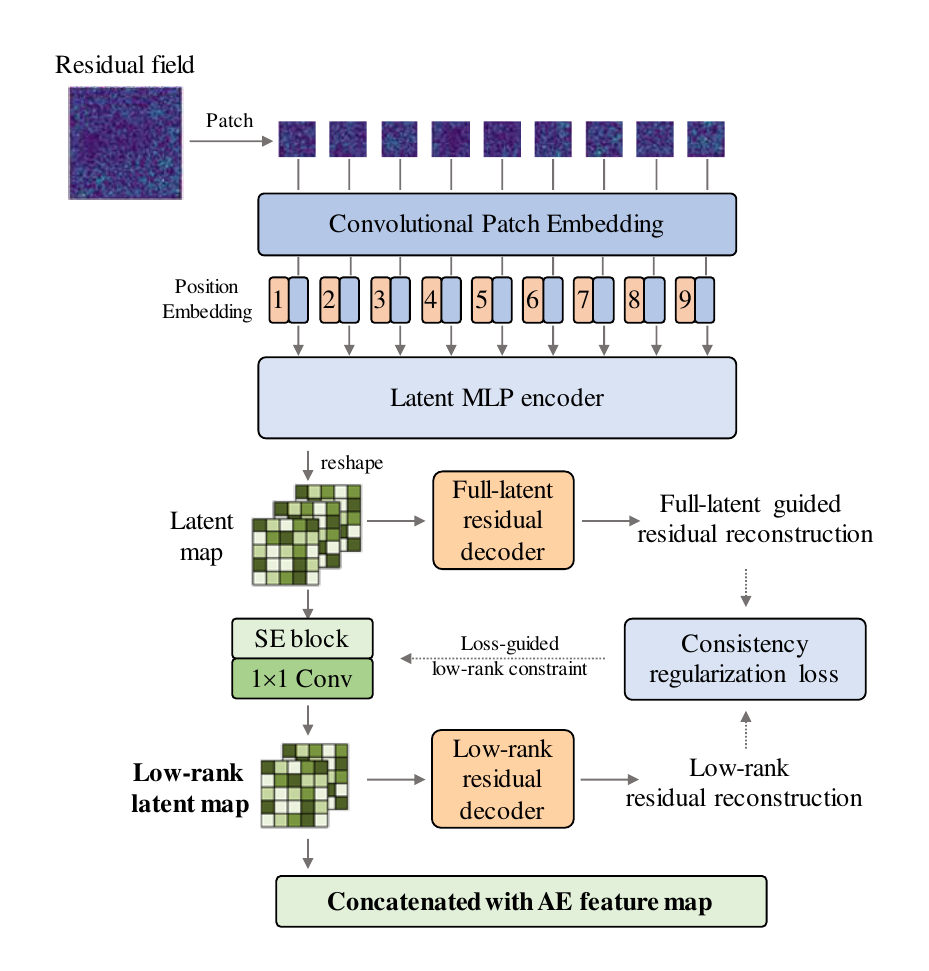}
	\caption{Overall architecture of LRC-FNO with latent-space and field-space residual closures.}
	\label{fig:fig2}
\end{figure}

\paragraph{}
The two-dimensional residual latent map is then flattened into a latent sequence and augmented with learnable positional latent embeddings:

\begin{equation}
	\label{equ:equ15}
	{{\bf{l}}_0} = {\rm{Flatten}}\left( {{{\bf{h}}_0}} \right) + {{\bf{e}}_{{\rm{pos}}}},\qquad {{\bf{l}}_0} \in {\mathbb{R}^{{N_l} \times {C_e}}}
\end{equation}

where ${N_l} = {H_l}{W_l}$ and ${{\bf{e}}_{{\rm{pos}}}} \in {\mathbb{R}^{{N_l} \times {C_e}}}$

\paragraph{}
The positional latent embedding preserves the spatial location information of each patch-wise latent element. The latent sequence is processed by the Latent MLP encoder:

\begin{equation}
	\label{equ:equ16}
	{\bf{l}} = {{\cal M}_\eta }\left( {{{\bf{l}}_0}} \right),\qquad {\bf{l}} \in {\mathbb{R}^{{N_l} \times {C_e}}}
\end{equation}

where $\mathcal{M}_{\eta}$ consists of normalization, linear projection, and nonlinear activation layers. The latent sequence is then reshaped back into a two-dimensional full residual latent map.

\begin{equation}
	\label{equ:equ17}
{{\bf{h}}_{{\rm{full}}}} = {\rm{Reshape}}\left( {\bf{l}} \right),\qquad {{\bf{h}}_{{\rm{full}}}} \in {^{{C_e} \times {H_l} \times {W_l}}}
\end{equation}

\paragraph{}
This full residual latent map preserves the local unresolved information extracted from the AE reconstruction residual while maintaining the spatial organization of the patch-wise latent representation.

\paragraph{}
To obtain a compact closure representation for the subsequent FNO solver, the Latent Closure Refiner further introduces a low-rank latent branch. The full residual latent map is first processed by an SE channel-reweighting module:

\begin{equation}
	\label{equ:equ18}
	{{\bf{h}}_{{\rm{SE}}}} = {\rm{SE}}\left( {{{\bf{h}}_{{\rm{full}}}}} \right)
\end{equation}

where the SE module\cite{cite40} emphasizes informative residual channels through global channel statistics and channel-wise reweighting. A 1×1 convolution is then used for channel reduction:

\begin{equation}
	\label{equ:equ19}
	{{\bf{z}}_{{\rm{LC}}}} = {{\cal R}_{1 \times 1}}\left( {{{\bf{h}}_{{\rm{SE}}}}} \right),\qquad {{\bf{z}}_{{\rm{LC}}}} \in {^{{C_{{\rm{LC}}}} \times {H_l} \times {W_l}}}
\end{equation}

where $\mathbf{z}_{\mathrm{LC}}$ denotes the low-rank latent map, $C_{\mathrm{LC}}$ is the number of low-rank closure channels, and $C_{\mathrm{LC}} < C_e$. This map is the compact residual closure representation passed to the downstream FNO field solver.

\paragraph{}
In the second training stage, the autoencoder and the Latent Closure Refiner are jointly optimized. The autoencoder is initialized from the first-stage reconstruction training and provides the dominant source reconstruction as well as the current residual estimate. The Latent Closure Refiner further learns the unresolved local information from this residual. This joint optimization allows the AE latent representation and the residual closure representation to adapt to each other, instead of learning a residual correction on a fixed AE manifold.

\paragraph{}
The Latent Closure Refiner is trained through two residual reconstruction paths. The first path uses the full residual latent map to generate a full-latent residual reconstruction:

\begin{equation}
	\label{equ:equ20}
\delta {{\bf{s}}_{{\rm{full}}}} = {{\cal D}_{{\rm{full}}}}\left( {{{\bf{h}}_{{\rm{full}}}}} \right)
\end{equation}

whereas the second path uses the low-rank latent map to generate a low-rank residual reconstruction:

\begin{equation}
	\label{equ:equ21}
	\delta {{\bf{s}}_{{\rm{lr}}}} = {{\cal D}_{{\rm{lr}}}}\left( {{{\bf{z}}_{{\rm{LC}}}}} \right)
\end{equation}

both reconstruction paths output residual correction fields with the same size as the source field
$\delta \mathbf{s}_{\mathrm{full}}, \delta \mathbf{s}_{\mathrm{lr}} \in \mathbb{R}^{C_s \times H \times W}$.

\paragraph{}
The corresponding residual-corrected source reconstructions are:

\begin{equation}
	\label{equ:equ22}
	{{\bf{\hat s}}_{{\rm{full}}}} = {{\bf{\hat s}}_{{\rm{AE}}}} + \delta {{\bf{s}}_{{\rm{full}}}}
\end{equation}
\begin{equation}
	\label{equ:equ23}
	{{\bf{\hat s}}_{{\rm{lr}}}} = {{\bf{\hat s}}_{{\rm{AE}}}} + \delta {{\bf{s}}_{{\rm{lr}}}}
\end{equation}

\paragraph{}
Since the autoencoder is still updated in the second stage, 
$r_{\mathrm{AE}}$ changes dynamically with the AE parameters. 
Therefore, the refiner is not trained against a fixed precomputed AE residual. 
Instead, the optimization is imposed on the final residual-corrected reconstructions:

\begin{equation}
	\label{equ:equ24}
{{\cal L}_{{\rm{LC}}}} = \parallel {{\bf{\hat s}}_{{\rm{AE}}}} + \delta {{\bf{s}}_{{\rm{full}}}} - {\bf{\tilde s}}\parallel _2^2 + \parallel {{\bf{\hat s}}_{{\rm{AE}}}} + \delta {{\bf{s}}_{{\rm{lr}}}} - {\bf{\tilde s}}\parallel _2^2
\end{equation}

\paragraph{}
The two reconstruction paths are coupled through the shared full residual latent map. 
Therefore, the full-latent path encourages $\mathbf{h}_{\mathrm{full}}$ to retain sufficient residual expressiveness, 
while the low-rank path forces compressed closure map $\mathbf{z}_{\mathrm{LC}}$ to preserve the essential residual information. 
This dual-path training introduces an implicit coupling between the full residual latent space and the low-rank closure latent space without requiring an additional output-consistency loss.

\paragraph{}
Finally, the AE feature map and the low-rank latent map are concatenated to form the closure-enhanced source representation:

\begin{equation}
	\label{equ:equ25}
{\bf{z}} = {\rm{Concat}}\left( {{{\bf{z}}_{{\rm{AE}}}},{{\bf{z}}_{{\rm{LC}}}}} \right),\qquad {\bf{z}} \in {\mathbb{R}^{({C_{{\rm{AE}}}} + {C_{{\rm{LC}}}}) \times {H_l} \times {W_l}}}
\end{equation}

\paragraph{}
This representation contains both the dominant source field features captured by the autoencoder and the unresolved residual features extracted by the Latent Closure Refiner. In this way, latent-space residual closure enriches the compressed source representation without directly feeding the full-resolution source field into the FNO solver, thereby improving the robustness and expressiveness of source field encoding.

\subsection{Field-space residual closure with Coarse-FNO Solver and Residual-Closure FNO}
\label{sec:sec3.2}

\paragraph{}
The second closure level of LRC-FNO is the field-space residual closure, which aims to learn a high-fidelity operator mapping from source fields to solution fields based on the latent source field representation. Instead of directly learning the complete solution field on the full-resolution grid, a two-stage FNO structure is adopted by combining coarse-scale prediction with full-resolution residual correction. The Coarse-FNO Solver in the first stage captures the dominant response from the source field to the solution field, while the Residual-Closure FNO in the second stage compensates for unresolved high-frequency structures, local gradients, and boundary-related residuals in the coarse solution.

\paragraph{}
After the latent-space closure module is trained, its encoded representation is used as the input of the field-space operator solver. 
To convert the low-dimensional latent representation $\mathbf{z}$ into a source field form that can be processed by FNO, 
a source proxy decoder $\mathcal{D}_{\mathrm{ecs}}$, composed of deconvolutional upsampling layers, is introduced to map $\mathbf{z}$ to a full-resolution source proxy:

\begin{equation}
	\label{equ:equ26}
	{\bf{\hat s}} = {\rm{De}}{{\rm{c}}_s}\left( {\bf{z}} \right),\qquad {\bf{\hat s}} \in {\mathbb{R}^{H \times W}}
\end{equation}

where, $\hat{\mathbf{s}}$ is not forced to reconstruct the original particle-deposited source field. 
Instead, it is optimized through the downstream field-prediction loss together with the FNO modules. 
Therefore, $\hat{\mathbf{s}}$ acts as a task-oriented source proxy for source-to-field operator learning. 
With this design, the FNO solver does not directly receive the original source field containing particle noise, 
but learns the field equation operator from the compressed source structure extracted by the autoencoder and the Latent Closure Refiner.

\paragraph{}
The FNO module used in this work can be written as a parallel combination of a spectral convolution branch, 
a pointwise linear weight branch, and a CNN local weight branch. 
Let $\mathbf{v}^{(\kappa)}$ denote the hidden feature at the $\kappa$-th layer. 
One FNO-CNN layer is expressed as:

\begin{equation}
	\label{equ:equ27}
	{{\bf{v}}^{(k + 1)}} = \sigma \left[ {{{\cal K}^{(k)}}\left( {{{\bf{v}}^{(k)}}} \right) + {{\cal W}^{(k)}}\left( {{{\bf{v}}^{(k)}}} \right) + {{\cal C}^{(k)}}\left( {{{\bf{v}}^{(k)}}} \right)} \right]
\end{equation}

where, $\mathcal{K}^{(\kappa)}$ denotes the Fourier spectral convolution, 
which models the global response from the source field to the solution field; 
$\mathcal{W}^{(\kappa)}$ denotes the pointwise linear weight mapping in the original FNO; 
and $\mathcal{C}^{(\kappa)}$ denotes the parallel CNN local weight branch, 
which enhances the representation of local neighborhood structures and boundary-related residuals. 
When the CNN local branch is not used, $\mathcal{C}^{(\kappa)}$ can be omitted, 
and the module reduces to the standard FNO form.

\paragraph{}
In the first stage, the Coarse-FNO Solver takes the source proxy $\hat{\mathbf{s}}$ as input, 
optionally concatenated with normalized coordinates $\mathbf{x}$ to enhance spatial awareness. 
It outputs the coarse-scale solution field on a downsampled grid:

\begin{equation}
	\label{equ:equ28}
	{{\bf{\hat u}}_{\rm{c}}} = {\rm{FN}}{{\rm{O}}_{\rm{c}}}\left( {{\bf{\hat s}},{\bf{x}}} \right),\qquad {{\bf{\hat u}}_{\rm{c}}} \in {^{{H_d} \times {W_d}}}
\end{equation}

where, $\mathbf{u}$ denotes the target physical field, such as the electrostatic potential $\phi$ or the magnetic vector potential $\mathbf{A}$, 
and $H_d$ and $W_d$ are the spatial sizes of the downsampled grid. 
The $\mathcal{S}_d$ denotes the downsampling operator, the supervised target for the first stage is

\begin{equation}
	\label{equ:equ29}
{{\bf{u}}_{\rm{d}}} = {{\cal S}_d}\left( {\bf{u}} \right)
\end{equation}

\paragraph{}
The training loss of the Coarse-FNO Solver is

\begin{equation}
{{\cal L}_{\rm{c}}} = \parallel {{\bf{\hat u}}_{\rm{c}}} - {{\bf{u}}_{\rm{d}}}\parallel _2^2
\end{equation}

\paragraph{}
The purpose of the Coarse-FNO Solver is not to directly produce the final full-resolution solution, but to learn the dominant physical response of the solution field. For PIC field-solving problems, the mapping from source fields to solution fields is usually highly nonlocal, such as the long-range electrostatic response in Poisson-type equations or the current-to-vector-potential mapping in magnetoquasistatic models. The spectral convolution in FNO can model such global responses in Fourier space, making it suitable as a coarse-scale operator learning module.

\paragraph{}
The Residual-Closure FNO in the second stage is used to recover full-resolution local details. First, the coarse-scale solution field is upsampled to the original grid to obtain the upsampled coarse solution:

\begin{equation}
{{\bf{\hat u}}_{{\rm{up}}}} = {{\cal U}_d}\left( {{{{\bf{\hat u}}}_{\rm{c}}}} \right),\qquad {{\bf{\hat u}}_{{\rm{up}}}} \in {\mathbb{R}^{H \times W}}
\end{equation}

where $\mathcal{U}_d$ denotes the upsampling operator. 
The upsampled coarse solution and the source proxy are then concatenated as the input $\mathbf{q}$ of the Residual-Closure FNO:

\begin{equation}
{\bf{q}} = {\rm{Concat}}\left( {{{{\bf{\hat u}}}_{{\rm{up}}}},{\bf{\hat s}}} \right)
\end{equation}

\paragraph{}
The Residual-Closure FNO predicts the solution field residual on the full-resolution grid:

\begin{equation}
\delta {{\bf{\hat u}}_{\rm{r}}} = {\rm{FN}}{{\rm{O}}_{\rm{r}}}\left( {{\bf{q}},{\bf{x}}} \right),\qquad \delta {{\bf{\hat u}}_{\rm{r}}} \in {^{H \times W}}
\end{equation}

\paragraph{}
The final full-resolution solution field is obtained by adding the coarse solution and the residual correction:

\begin{equation}
{\bf{\hat u}} = {{\bf{\hat u}}_{{\rm{up}}}} + \delta {{\bf{\hat u}}_{\rm{r}}}
\end{equation}

\paragraph{}
The training loss of the second stage is:

\begin{equation}
{{\cal L}_{\rm{r}}} = \parallel {{\bf{\hat u}}_{{\rm{up}}}} + \delta {{\bf{\hat u}}_{\rm{r}}} - {\bf{u}}\parallel _2^2
\end{equation}

\paragraph{}
The residual in this stage is different from the AE source field reconstruction residual defined in Section 3.1. It represents an operator-induced residual in the solution field, arising from the unresolved errors of the coarse source-to-field mapping. The Coarse-FNO Solver captures the dominant source-to-field response, whereas the Residual-Closure FNO learns a full-resolution error closure associated with downsampling, spectral truncation, and the loss of local details.

\paragraph{}
Therefore, the proposed neural field solver does not directly predict the full-resolution solution field from the latent variables in a single step. Instead, high-fidelity field prediction is achieved through a cascaded structure consisting of source-proxy decoding, coarse-scale FNO solving, and full-resolution residual closure. Complementary to the latent-space residual closure introduced in Section 3.1, the field-space residual closure in Section 3.2 further compensates for unresolved errors in the learned source-to-field operator mapping, thereby forming the two-level residual-closure framework of LRC-FNO.

\section{Numerical Experiments}
\label{sec:sec4}

\paragraph{}
To evaluate the performance of the proposed LRC-FNO framework, numerical experiments are conducted on three representative PIC field-solving problems, including 1D linear Landau damping (LLD), 2D two-stream instability (TSI), and a 2D scrape-off layer (SOL) plasma model. These cases cover different physical mechanisms, spatial dimensions, source field characteristics, and boundary-related responses, thereby providing a progressive validation of the proposed method from canonical kinetic benchmarks to more complex plasma edge simulations.

\paragraph{}
For each case, the conventional numerical field solver is used as the reference solver to generate training and testing data. The proposed LRC-FNO is then trained to replace the field-solving step by learning the mapping from particle-deposited source fields to the corresponding solution fields. In all cases, the Coarse-FNO Solver predicts the solution field on a coarse grid with one-quarter of the original spatial resolution, while the Residual-Closure FNO further learns the full-resolution residual correction. All source and solution fields used for neural-network training are standardized using the statistics of the training dataset to ensure consistent scaling across different physical quantities and simulation cases. The training and test sets are separated by physical parameters or random seeds rather than by random time-step sampling within the same trajectory.

\paragraph{}
All neural networks are implemented using the PyTorch framework\cite{cite41} and optimized with the Adam optimizer\cite{cite42}. 
The learning rates are fixed at $1\times10^{-4}$. 
All the numerical experiments are conducted on Intel\textsuperscript{\textregistered} Core i9-9900K @ 3.60\,GHz / NVIDIA GeForce RTX 3090. 
The following subsections present the dataset construction, training settings, and prediction performance for the three PIC benchmarks. 
The linear Landau damping case is first used to examine whether LRC-FNO can reproduce wave damping dynamics in a one-dimensional electrostatic system. 
The two-stream instability case further evaluates the model in a two-dimensional nonlinear instability problem with vortex-like phase-space evolution. 
Finally, the scrape-off layer case tests the applicability of LRC-FNO to a more complex plasma edge configuration involving sheath-related structures and boundary-sensitive field responses.

\subsection{1D Linear Landau Damping}
\label{sec:sec4.1}

\paragraph{}
The first benchmark is a 1D linear Landau damping (LLD) problem, following the benchmark in Ref.\cite{cite9}. This case is used to evaluate the ability of LRC-FNO to learn the field-solving operator in a standard electrostatic PIC system. The model is based on the 1D Vlasov-Poisson plasma system, where electrons evolve self-consistently under the electrostatic field, while ions are treated as a fixed uniform neutralizing background. Since Landau damping involves typical wave-particle interaction, the amplitude decay and phase evolution of the electrostatic potential mode are sensitive to the accuracy of the field solver. Therefore, it provides a suitable benchmark for validating the surrogate field solver.

\paragraph{}
In this case, the electrostatic field is governed by the 1D Poisson equation. 
The computational domain uses periodic boundary conditions, with the domain length set to 
$L = 100\,\lambda_D$, where $\lambda_D$ is the Debye length. 
The spatial grid contains $N_x = 512$ cells, and the particles-per-cell $\mathrm{PPC} = 2000$. 
The time step is set to $\Delta t = 0.1\omega_{pe}^{-1}$, 
and the simulation is advanced to $\omega_{pe}t = 100$. 
The initial particle positions are uniformly distributed and then perturbed by a prescribed single-mode density perturbation to excite a Langmuir wave. 
The initial electron velocities are sampled from a zero-drift Maxwellian distribution. 
At each time step, the electron charge density is first deposited onto the grid and smoothed. 
The reference electrostatic potential and electric field are then obtained using an FFT-based Poisson solver, 
after which the electric field is interpolated to particle positions for particle pushing.

\paragraph{}
A single-mode density perturbation is imposed initially. 
In this work, the mode number is fixed as $M = 4$, with the corresponding wavenumber 
$k_M = \dfrac{2\pi M}{L}$. 
For a given perturbation amplitude $A$, the initial particle positions are perturbed as:

\begin{equation}
{x_p}(0) = {x_{p,0}} + \frac{A}{{{k_M}}}\sin \left( {{k_M}{x_{p,0}}} \right)
\end{equation}

where $x_{p,0}$ denotes the uniformly distributed initial particle position. 
By varying the perturbation amplitude $A$, Landau damping dynamics under different perturbation strengths can be obtained.

\paragraph{}
The LLD dataset is constructed by scanning the perturbation amplitude 
$A \in [0.004, 0.005, 0.01, 0.03, 0.1, 0.2, 0.25, 0.4, 0.5]$. 
For each amplitude, the charge density and electrostatic potential are recorded at all time steps to form source-solution pairs. 
The learning task of LRC-FNO in this benchmark is the instantaneous electrostatic field-solving mapping. 
To evaluate the generalization capability of the model across different perturbation amplitudes, 
the cases $A \in [0.004, 0.2, 0.5]$ are selected as the test set, 
while the remaining amplitudes are used for training. 
This amplitude-based partition covers Langmuir-wave evolution from weak to finite-amplitude perturbations, 
allowing us to examine whether the model can preserve the amplitude damping and phase evolution of the electrostatic potential mode under different perturbation strengths, 
rather than merely fitting temporally adjacent snapshots.

\begin{figure}[!b]
	\centering
	\includegraphics[width=17cm]{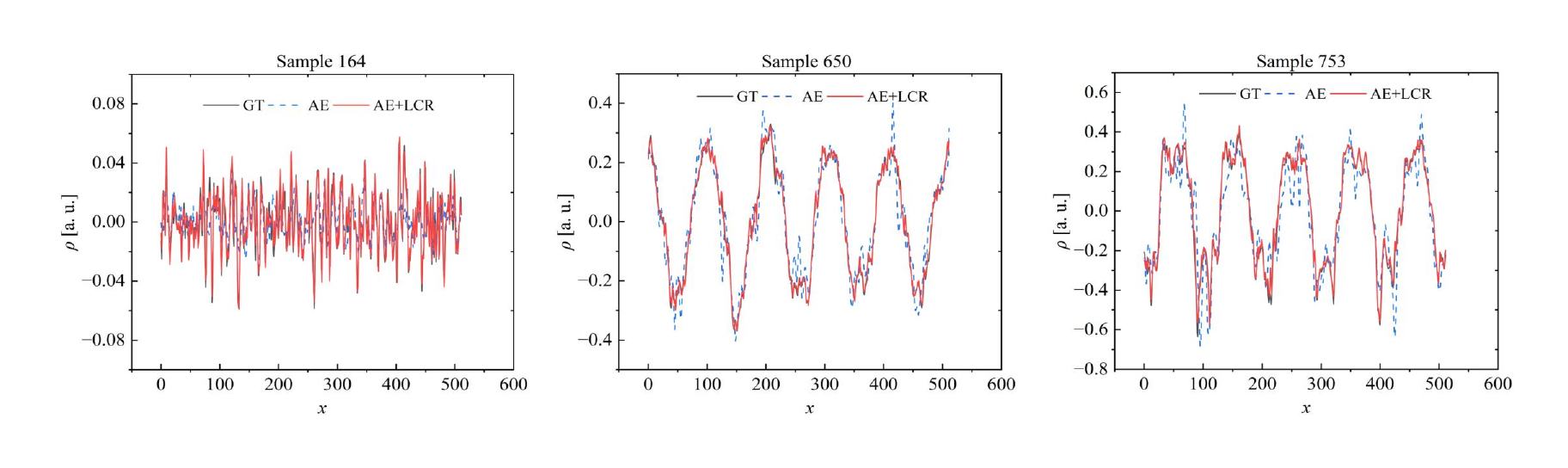}
	\caption{Comparison of charge density reconstruction by AE and AE+LCR in the 1D LLD benchmark.}
	\label{fig:fig2}
\end{figure}

\paragraph{}
Figure~4 shows three randomly selected charge density $\rho$ reconstruction samples from the test set, 
comparing the source field reconstruction capability of the autoencoder alone and the autoencoder with the Latent Closure Refiner (AE+LCR). 
The average relative $L_2$ error of AE on the test set is 0.0969, while the error is reduced to 0.0195 after introducing LCR. 
This demonstrates that the latent-space residual closure substantially improves the reconstruction accuracy of the source field.

\paragraph{}
As shown in Figure~4, the charge density $\rho$ fields in the LLD case contain pronounced local oscillations and particle-noise-induced structures. 
When only AE is used, part of the local information is lost during compression, leading to noticeable discrepancies in peak values and local phase variations. 
In contrast, AE+LCR further compensates for the local residual information based on the dominant low-dimensional structure learned by AE, 
thereby recovering the variations of the charge density more accurately. 
This indicates that the Latent Closure Refiner does not merely increase the network complexity, 
but provides a targeted closure correction for the truncation error introduced by AE compression.

\paragraph{}
For ablation comparison, two latent neural-operator baselines are constructed. The first baseline is denoted as L-FNO, namely Latent FNO, which directly takes the compressed latent representation as input and predicts the full-resolution solution field using a single-stage FNO. The second baseline is denoted as L-FNO-CNN, namely Latent FNO-CNN, which further introduces a parallel CNN local branch into the FNO module to enhance the representation of local structures. The proposed method is denoted as LRC-FNO, namely Latent Residual-Closure FNO, which incorporates both latent-space residual closure and field-space residual closure through the Latent Closure Refiner, Coarse-FNO Solver, and Residual-Closure FNO. All three models use the low-dimensional latent representation z generated by AE+LCR as the input. This ablation is designed to isolate the effect of the field-space residual closure, since all three models use the same AE+LCR latent representation as input.

\paragraph{}
After training, the average relative $L_2$ errors of L-FNO, L-FNO-CNN, and LRC-FNO on the test set are 
0.0542, 0.0177, and 0.0180, respectively. 
The results show that introducing either the CNN local branch or the residual closure structure significantly reduces the overall prediction error compared with the L-FNO. 
This indicates that a single-stage spectral operator mapping from latent variables to the solution field is insufficient to fully recover the local structures of the electrostatic potential. 
Although L-FNO-CNN achieves a single-step field error comparable to that of LRC-FNO, 
their long-term dynamical behaviors become clearly different after being embedded into the closed-loop PIC time integration.

\paragraph{}
After replacing the FFT-based Poisson solver in the original PIC framework with the three surrogate solvers, 
the complete evolution results under the test conditions 
$A \in \{0.004, 0.2, 0.5\}$ are shown in Figure~5. 
Figures~5(a)--(c) present the full time evolution of 
$\left|\operatorname{Re}\hat{\phi}_4\right|$ under different perturbation amplitudes, 
while Figures~5(d)--(f) show enlarged views of the late-stage evolution. 
Here, $\left|\operatorname{Re}\hat{\phi}_4\right|$ denotes the fourth spatial Fourier coefficient of the electrostatic potential, 
corresponding to the initially excited mode $M = 4$. 
Therefore, $\left|\operatorname{Re}\hat{\phi}_4\right|$ characterizes the modal-amplitude evolution of the excited Langmuir wave. 
Its oscillation frequency, amplitude decay, and phase variation directly reflect the wave-particle interaction features of the Landau damping process. 
This diagnostic not only evaluates the accuracy of single-step potential prediction, 
but also examines whether the surrogate field solver can preserve the correct long-term dynamics after being embedded into closed-loop PIC time integration.

\begin{figure}[!b]
	\centering
	\includegraphics[width=17cm]{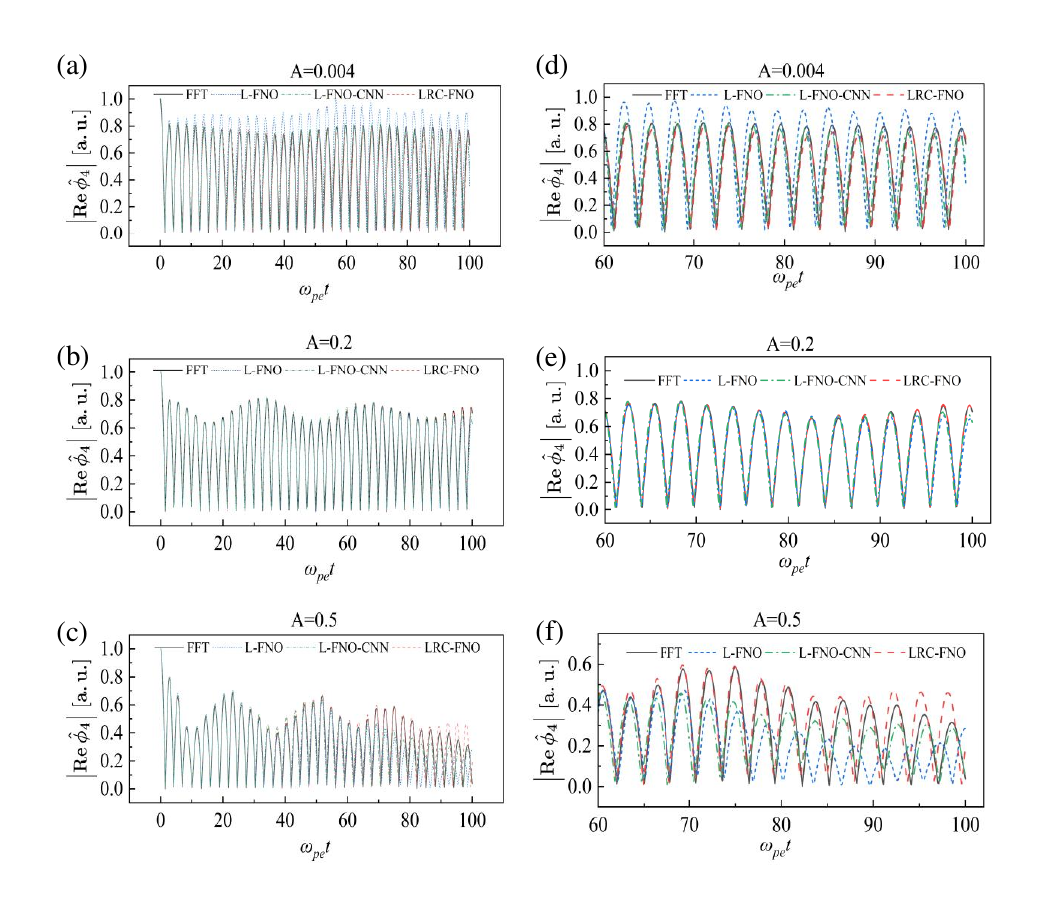}
	\caption{Comparison of the fourth electrostatic potential Fourier mode evolution using FFT and surrogate field solvers in the 1D LLD benchmark.}
	\label{fig:fig2}
\end{figure}

\paragraph{}
In the dataset split, $A = 0.004$ and $A = 0.5$ correspond to low-amplitude and high-amplitude extrapolation tests, respectively, 
whereas $A = 0.2$ lies within the training amplitude range and is regarded as an interpolation test. 
Therefore, all three models show their best overall performance at $A = 0.2$.

\paragraph{}
Comparing different models, LRC-FNO achieves the best agreement with the FFT reference solution in both amplitude and phase at $A = 0.2$, 
and it can more accurately reproduce the oscillatory damping process of LLD. 
In contrast, although L-FNO and L-FNO-CNN can capture the dominant oscillation frequency, 
they still exhibit amplitude drift in the later stage. 
Under the low-amplitude extrapolation case $A = 0.004$, L-FNO deviates from the reference trajectory at an early stage, 
whereas L-FNO-CNN and LRC-FNO maintain more robust oscillation patterns, 
indicating that the CNN local branch contributes positively to local error compensation.
Under the high-amplitude extrapolation case $A = 0.5$, the perturbation is stronger, 
and the spatial variations of charge density and electrostatic potential become more intense. 
The closed-loop PIC evolution is therefore more sensitive to field-solver errors. 
In this case, all three solvers show amplitude and phase deviations to some extent, 
but LRC-FNO remains the closest to the FFT reference solution, 
demonstrating better long-term dynamical preservation.

\begin{table}[htbp]
	\centering
	\caption{Quantitative comparison of modal-amplitude evolution in the 1D LLD benchmark.}
	\label{tab:lld_modal_amplitude}
	\begin{tabular}{lcccccc}
		\toprule
		& \multicolumn{2}{c}{$A = 0.004$}
		& \multicolumn{2}{c}{$A = 0.2$}
		& \multicolumn{2}{c}{$A = 0.5$} \\
		
		& $R^2$ & $\mathrm{NCC}_{\max}$
		& $R^2$ & $\mathrm{NCC}_{\max}$
		& $R^2$ & $\mathrm{NCC}_{\max}$ \\
		\midrule
		L-FNO     & 0.379 & 0.924 & 0.992 & 0.997 & 0.623 & 0.841 \\
		L-FNO-CNN & 0.911 & 0.981 & 0.971 & 0.986 & 0.848 & 0.950 \\
		LRC-FNO   & 0.957 & 0.992 & 0.999 & 0.999 & 0.917 & 0.962 \\
		\bottomrule
	\end{tabular}
\end{table}

\paragraph{}
Table~1 shows the $R^2$ and $\mathrm{NCC}_{\max}$ of the three models under different test conditions. 
Here, $R^2$ evaluates the overall agreement between the predicted modal-amplitude sequence and the FFT reference sequence, 
while $\mathrm{NCC}_{\max}$, the maximum normalized cross-correlation coefficient, measures waveform similarity and phase consistency under allowable time lag. 
As shown in the table, LRC-FNO achieves the highest $R^2$ and $\mathrm{NCC}_{\max}$ values for all three test amplitudes. 
The improvement is especially evident in the extrapolation cases $A = 0.004$ and $A = 0.5$, 
showing that LRC-FNO not only reduces single-step field prediction error but also better preserves modal-amplitude evolution and phase relationships in closed-loop PIC simulations. 
These results indicate that the two-level residual closure structure of LRC-FNO effectively compensates for spatial information loss caused by latent compression and spectral truncation, 
thereby improving the robustness and physical consistency of the source-to-field mapping in long-time wave-particle interaction simulations.

\subsection{2D Two-Stream Instability}
\label{sec:sec4.2}

\paragraph{}
The second benchmark considers the 2D two-stream instability (TSI) problem to evaluate the field-solving capability of LRC-FNO in a 2D electrostatic PIC system\cite{10}. Compared with the 1D LLD case, the 2D TSI problem simultaneously involves beam-driven instability growth along the streaming direction, transverse mode coupling, and the evolution of 2D spatial structures, thereby imposing stricter requirements on the spatial generalization capability and long-term closed-loop consistency of the surrogate solver.

\paragraph{}
This benchmark is based on the 2D Vlasov--Poisson model. Electrons move inside the periodic domain $[0,L_x]\times[0,L_y]$, while ions are treated as a fixed uniform neutralizing background. The system satisfies the 2D Poisson equation and employs 2D particle deposition. After depositing the charge density onto the grid, the electrostatic potential is solved using an FFT-based Poisson solver in Fourier space, and the electric field is calculated through spectral derivatives. The electric field is then interpolated back to particle positions for particle pushing. The computational domain sizes are chosen as $L_x = 100$ and $L_y = 50$, with grid resolutions $N_x = 128$ and $N_y = 64$. Periodic boundary conditions are imposed in both spatial directions, and the $\mathrm{PPC} = 50$. The initial electrons are spatially uniform and consist of two counter-streaming Maxwellian electron beams drifting along the $x$ direction. The two electron beams have drift velocities $v_0 = \pm 3$. The simulation time step is $\Delta t = 0.1$, and the total simulation time is $T = 30$.

\paragraph{}
At each time step, the 2D charge density field and electrostatic potential field are recorded to form source-solution pairs. The dataset is constructed by varying the random initialization seed, resulting in 100 evolution samples with different particle initializations. Different random seeds correspond to different particle-noise perturbations and initial microscopic fluctuations, leading to different instability growth processes and two-dimensional field-structure evolution. The dataset is further divided according to the random seeds, where 80 samples are used for training and the remaining 20 samples are used for testing.

\begin{figure}[!b]
	\centering
	\includegraphics[width=15cm]{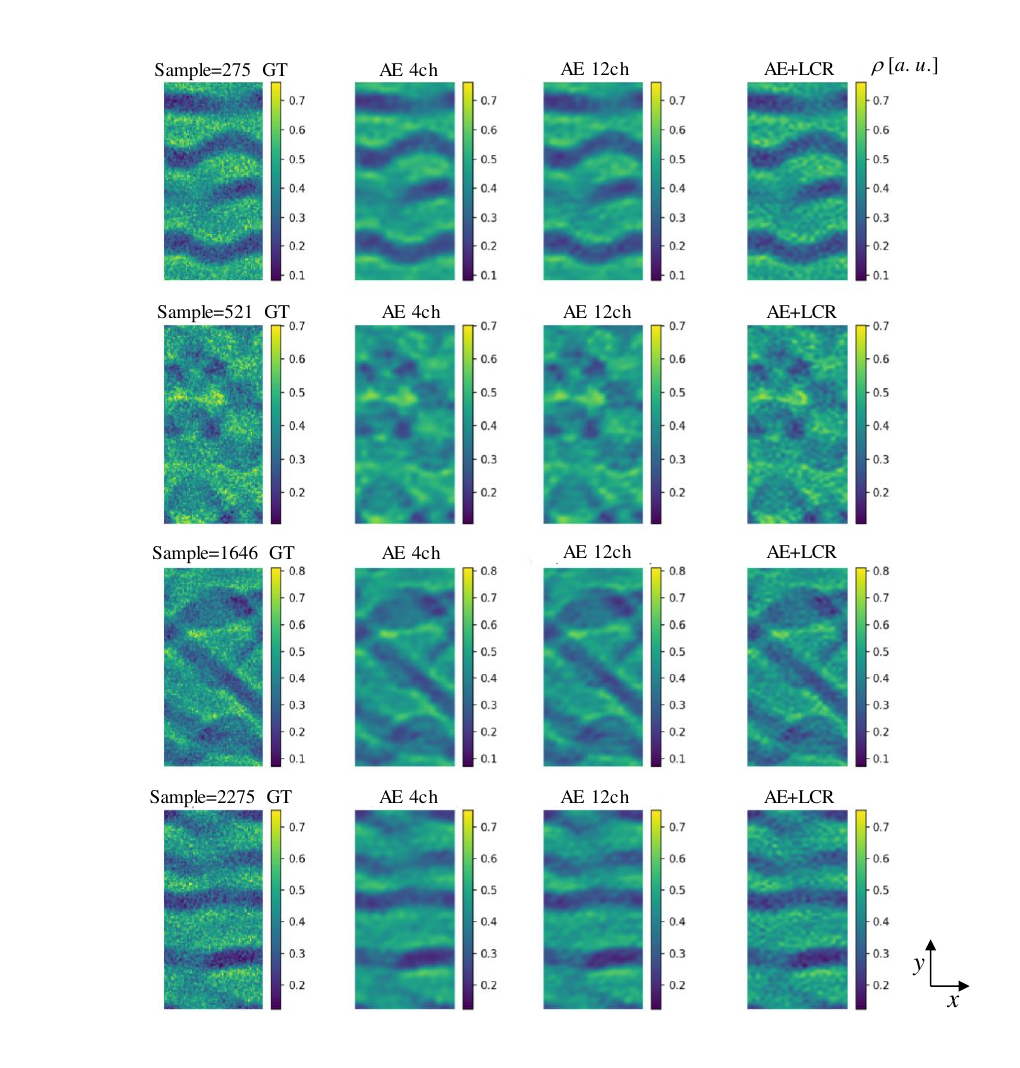}
	\caption{Comparison of charge density reconstruction using AE-4ch, AE-12ch, and AE+LCR in the 2D TSI benchmark.}
	\label{fig:fig6}
\end{figure}

\paragraph{}
In this benchmark, an ablation study is further conducted to evaluate the effectiveness of the latent-space residual closure. 
Specifically, the latent representation $\mathbf{z}$ used by the Coarse-FNO Solver and the Residual-Closure FNO is constructed from three different source-field compression strategies: 
AE-4ch, which uses only an autoencoder with latent dimension $C_{\mathrm{AE}} = 4$; 
AE-12ch, which uses only an autoencoder with latent dimension $C_{\mathrm{AE}} = 12$; 
and the proposed AE+LCR framework. 
In AE+LCR, the autoencoder latent feature dimension is 4 channels, while the residual latent map dimension is 8 channels, resulting in a total latent dimension of 12 channels. 
Different from the LLD ablation, this test focuses on the effect of latent-space source-field representation. 
The same two-stage field-space solver is used for all three cases, while only the latent representation is changed.

\paragraph{}
The relative $L_2$ errors of the three source-field reconstruction methods on the test set are 0.130, 0.128, and 0.116, respectively. 
The reconstructed charge density fields for randomly selected testing samples are shown in Figure~6. 
From the relative $L_2$ errors, the differences among the three methods are relatively small. 
However, significant differences can be observed in the reconstructed spatial structures. 
When only the autoencoder is used, the reconstructed source fields exhibit strong smoothing effects, and many local high-frequency structures and fine-scale fluctuations are substantially suppressed. 
Even when the latent dimension is increased to 12 channels, this smoothing behavior still remains.

\paragraph{}
In contrast, AE+LCR is able to recover substantially richer local details while maintaining a low-dimensional latent representation. 
In particular, for the local stripe structures, oblique wave patterns, and particle-noise-dominated fine-scale regions generated during the evolution of the two-stream instability, 
the AE+LCR reconstruction is noticeably closer to the ground truth. 
This indicates that the Latent Closure Refiner successfully learns the residual structures truncated during AE compression and compensates for unresolved local information through latent-space residual closure.

\paragraph{}
The latent representations $\mathbf{z}$ produced by AE-4ch, AE-12ch, and AE+LCR are further used as inputs to train the two-stage FNO solver, respectively. 
The resulting operator models achieve relative $L_2$ errors of 0.0390, 0.0442, and 0.0346 on the test set. 
These results indicate that the latent representation compressed by AE+LCR provides more informative features for the subsequent FNO training, thereby leading to higher accuracy.

\begin{figure}[!b]
	\centering
	\includegraphics[width=10cm]{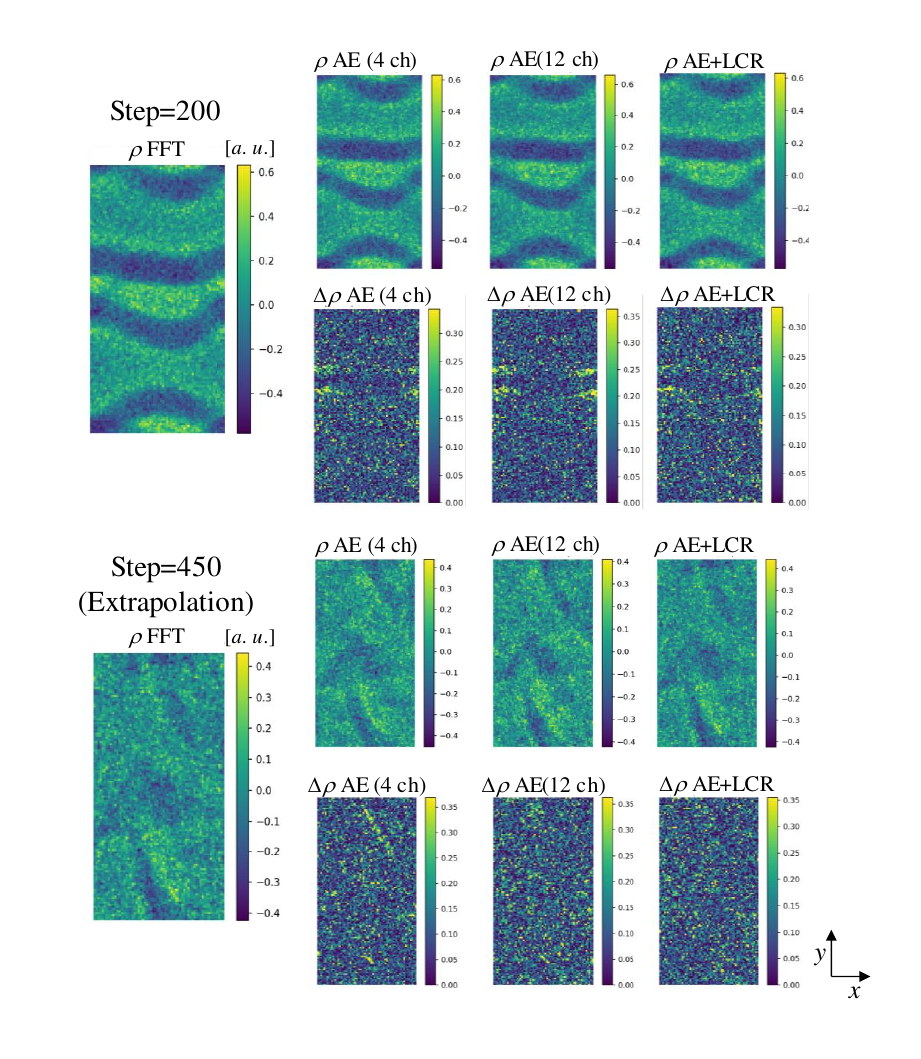}
	\caption{Charge density distributions and errors in closed-loop 2D TSI simulations using different latent representations.}
	\label{fig:fig7}
\end{figure}

\begin{figure}[!b]
	\centering
	\includegraphics[width=15cm]{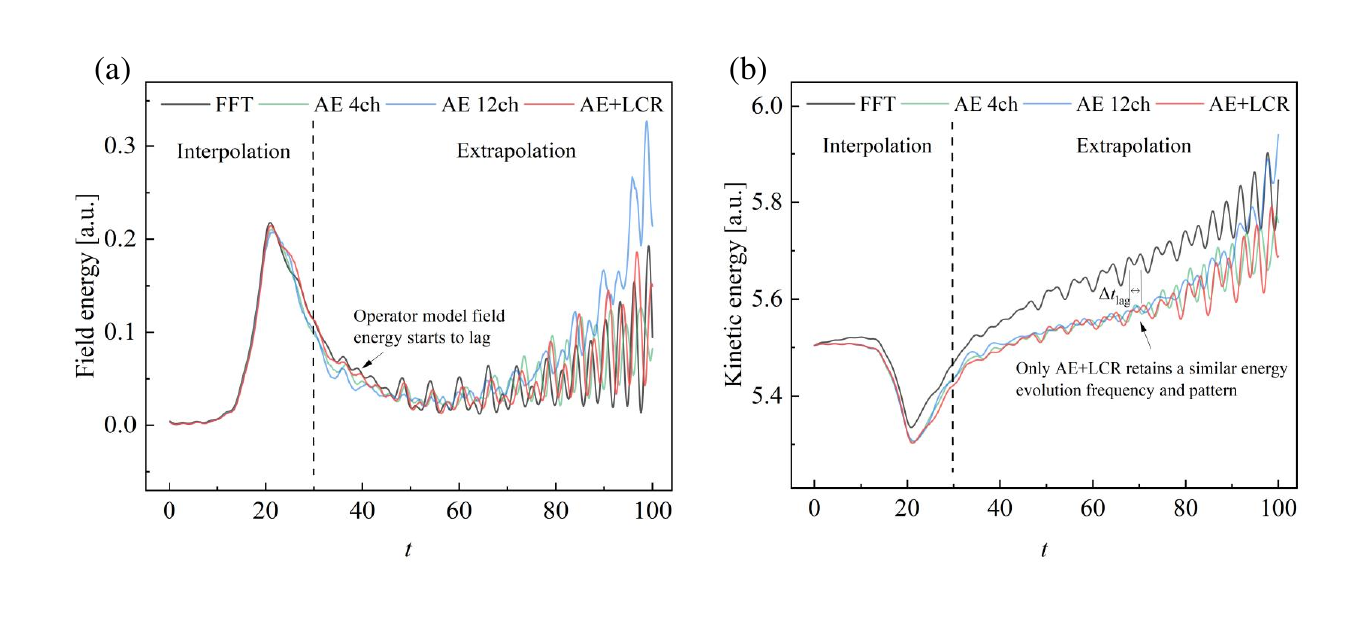}
	\caption{Comparison of field-energy and kinetic-energy evolution using different latent representations in closed-loop 2D TSI simulations.}
	\label{fig:fig8}
\end{figure}

\paragraph{}
To further evaluate their performance in closed-loop simulations, the three trained operator models are embedded into the PIC framework and evolved using random seeds that are not included in the training set. 
The charge-density distributions at $\mathrm{step} = 200$ and $\mathrm{step} = 450$ are shown in Figure~7. 
As seen in Figure~7, at $\mathrm{step} = 200$, the system is still within the main evolution regime covered by the training data. 
It should be emphasized that this does not mean the source field itself belongs to the training set, but rather that its dynamical state still lies within a similar evolution manifold. 
At this stage, the operator models based only on AE-compressed inputs already exhibit more pronounced error-concentrated regions in the charge density field than AE+LCR, indicating stronger structural deviations. 
In contrast, the AE+LCR-based model remains much closer to the FFT reference solution and better preserves the local structural features.

\paragraph{}
As the time evolution proceeds to the extrapolated $\mathrm{step} = 450$, the differences among the three models become more evident. 
At this stage, the operator model based on AE+LCR still does not show significant structural errors, whereas the structural errors of AE-12ch remain noticeable, and those of AE-4ch become substantially more severe. 
This suggests that even though the subsequent two-stage FNO introduces field-space residual closure, if the input latent representation itself does not preserve sufficient source field information, the error can still accumulate during the PIC evolution. 
Such accumulated field errors are then fed back into particle motion, eventually leading to nonphysical charge accumulation or distorted density structures.

\paragraph{}
Figure 8 shows the temporal evolution of the field energy and particle kinetic energy after embedding the three surrogate solvers into the closed-loop PIC simulation. The region before the dashed line corresponds to the time range in training data evolution, while the region after the dashed line represents the extrapolation regime. In the early stage, all three surrogate solvers can generally follow the FFT reference solution, and the overall trends of both field energy and kinetic energy remain consistent, indicating that the three models can provide reasonable field predictions near the training distribution.

\paragraph{}
As the simulation enters the extrapolation regime, the differences between the surrogate solvers and the FFT reference become more pronounced. For the field energy evolution shown in Figure 8(a), the operator models based on AE-compressed inputs begin to exhibit clear phase lag and amplitude deviations. In contrast, the operator model based on AE+LCR demonstrates better extrapolation stability, with the smallest phase lag and an oscillation pattern closer to the FFT reference. This indicates that the latent-space residual closure provides more complete source field information for the subsequent FNO solver and helps reduce the accumulation of field-solving errors during extrapolation.

\paragraph{}
A similar trend can also be observed in the particle kinetic energy, which reflects the collective response of particles to the electric field. As shown in Figure 8(b), the kinetic energy curve driven by AE+LCR remains the closest to the FFT reference in the extrapolation regime. It preserves not only the overall energy-growth trend but also a similar oscillation frequency and pattern. By contrast, the solvers based only on AE-compressed inputs gradually develop larger phase differences and pattern deviations, suggesting that the field solving errors have been fed back into particle motion and have further affected the particle dynamics.

\paragraph{}
Therefore, Figure 8 further demonstrates that the single-step field prediction error alone is insufficient to fully evaluate the performance of a surrogate field solver in closed-loop PIC simulations. Compared with operator models relying only on AE-compressed latent variables, the latent-space residual closure formed by AE+LCR more effectively preserves local source field structures and reduces the accumulated feedback of field errors into particle dynamics during long-time integration. This indicates that, in the TSI benchmark, LRC-FNO is able not only to predict the instantaneous electrostatic potential field, but also to reproduce the key kinetic characteristics observed in the reference PIC simulation during closed-loop evolution, including particle-field energy exchange, energy-oscillation frequency, and the coupled particle-field response. These results show that LRC-FNO achieves not only high single-step accuracy, but also improved physical consistency and extrapolation robustness in long-term evolution, making it more suitable for surrogate field solving in PIC simulations.

\subsection{2D Scrape-Off Layer Model}
\label{sec:sec4.3}
\paragraph{}
The third benchmark is a 2D scrape-off layer model (SOL), which is used to evaluate the surrogate field-solving capability of LRC-FNO under complex boundary conditions and in a more practical plasma scenario. 
This benchmark is constructed by referring to the 2D SOL PIC model of Cui et al.\cite{cite7}, while a simplified model is adopted in this work for validating surrogate field solving. 
The computational domain is a 2D rectangular region in $(x,y)$, and the particle velocities retain three components, corresponding to a 2D3V PIC model. 
The computational domain sizes are chosen as $L_x = 50\,\lambda_D$ and $L_y = 150\,\lambda_D$, with grid resolutions $N_x = 100$ and $N_y = 300$. 
Periodic boundary conditions are imposed in the $y$ direction, while wall boundaries are applied in the $x$ direction. 
As shown in Figure~9, a biased target segment is placed at the middle of the left wall, and the remaining wall regions are grounded. 
The external magnetic field forms a prescribed angle $\theta$ with the wall normal to describe particle transport under an oblique magnetic field.
\begin{figure}[!b]
	\centering
	\includegraphics[width=5cm]{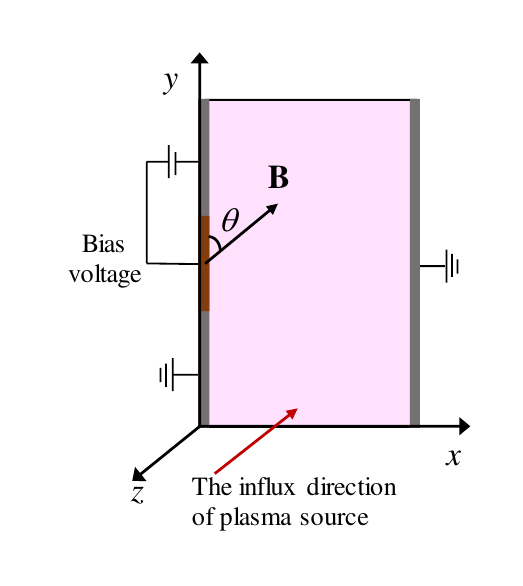}
	\caption{Schematic of the 2D SOL model with boundary conditions and oblique magnetic field.}
	\label{fig:fig9}
\end{figure}

\paragraph{}
Unlike the periodic Poisson problems in LLD and TSI, the electrostatic potential in this benchmark is constrained by both the biased boundary and grounded wall boundaries. Therefore, the reference solution cannot be obtained using an FFT-based Poisson solver, but must be computed using finite-difference discretization and an iterative solver. This makes the benchmark more suitable for demonstrating the computational cost of complex-boundary PIC field solving and the acceleration value of surrogate field solvers. To separate the boundary-driven potential from the self-consistent plasma response, the total electrostatic potential is decomposed into a Laplace boundary solution and a self-consistent plasma potential:

\begin{equation}
\phi  = {\phi _l} + {\phi _c}
\end{equation}

\paragraph{}
where $\phi_l$ is determined by the fixed biased and grounded boundary conditions and can be precomputed once at the beginning of the simulation. 
The self-consistent component $\phi_c$ is driven by the instantaneous charge density and must be updated at each time step.

\paragraph{}
Therefore, the electrostatic learning task of LRC-FNO is $\rho \rightarrow \phi_c$. 
This process can be viewed as learning a Green's-function-like response under complex boundary conditions, 
namely a nonlocal mapping from the charge source to the self-consistent potential response under fixed geometry and boundary constraints.

\paragraph{}
In the present 2D3V configuration, the dominant self-magnetic response considered in this work is driven by the out-of-plane current $J_z$. 
Therefore, the vector potential is represented by its out-of-plane component $A_z$. 
The corresponding learning task is $J_z \rightarrow A_z$. 
Thus, the SOL benchmark contains two surrogate field-solving tasks: 
mapping charge density to the self-consistent potential and mapping out-of-plane current density to the magnetic vector potential.

\paragraph{}
In this benchmark, the biased target voltage is fixed at $V_{\mathrm{bias}} = 12\,\mathrm{V}$, 
and the external magnetic field strength is set to $B_{\mathrm{ext}} = 1.42\,\mathrm{T}$. 
A parametric scan is performed over the biased-target width and the incident angle of the external magnetic field. 
The normalized biased-target width is set to 2, 4, 6, and 8, while the magnetic field angle is set to 
$\theta = 30^{\circ}, 45^{\circ}, 60^{\circ},$ and $90^{\circ}$. 
For each parameter case, 500 time steps are simulated, and the resulting parameter space is divided into training and test sets with a ratio of $6{:}4$.

\paragraph{}
After training, the relative $L_2$ reconstruction errors of AE+LCR for the charge density $\rho$ and the current density $J_z$ on the test set are 0.125 and 0.126, respectively. 
The relative $L_2$ prediction errors of LRC-FNO for the self-consistent potential $\phi_c$ and the magnetic vector potential $A_z$ are 0.0447 and 0.0251, respectively. 
These results indicate that, in single-step field prediction, LRC-FNO maintains sufficient prediction accuracy for both $\phi_c$ and $A_z$, providing a basis for subsequent closed-loop PIC integration.

\paragraph{}
However, when LRC-FNO is directly embedded into the closed-loop PIC evolution, the charge density distribution at the prescribed simulation step = 500 remains globally close to the reference solution, but its relative L2 error reaches approximately 0.1, as shown in Figure 10. The error distribution shows that the charge density error is mainly concentrated near the sheath region, accompanied by stripe-like structures along the y direction. This indicates that, in the SOL benchmark with mixed boundary constraints and strong localized sheath responses, the particle dynamics become more sensitive to field solving errors. Although LRC-FNO can provide a fast and reasonably accurate field solution, local field errors are first excited near the sheath and then further propagate along the y direction under the combined effects of periodic boundaries and oblique magnetic-field transport, leading to stripe-like error patterns.

\begin{figure}[!b]
	\centering
	\includegraphics[width=15cm]{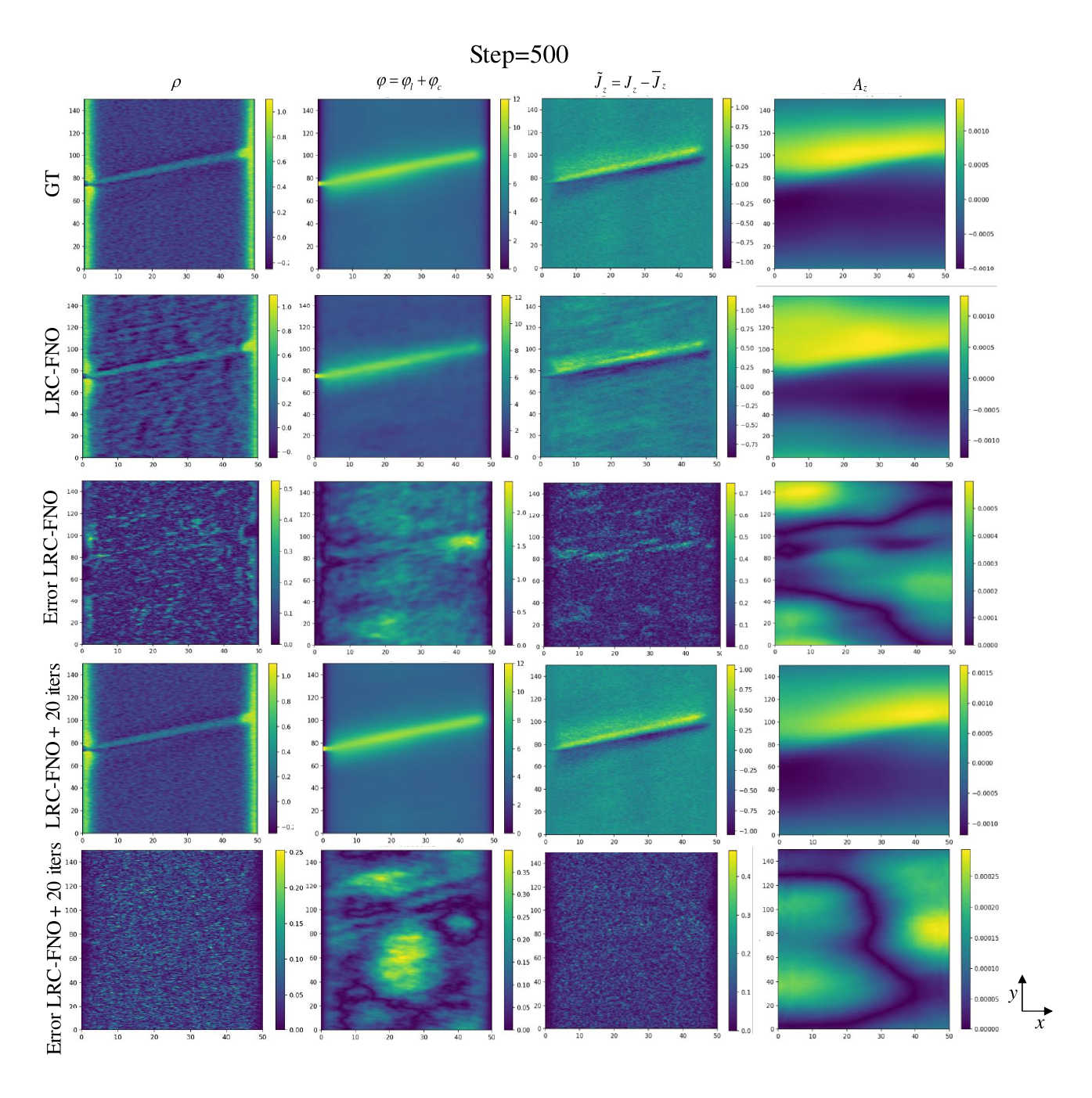}
	\caption{Multi-field distributions and errors at step = 500 in the SOL-PIC benchmark using direct LRC-FNO and LRC-FNO-assisted iterative correction.}
	\label{fig:fig10}
\end{figure}

\begin{figure}[!b]
	\centering
	\includegraphics[width=15cm]{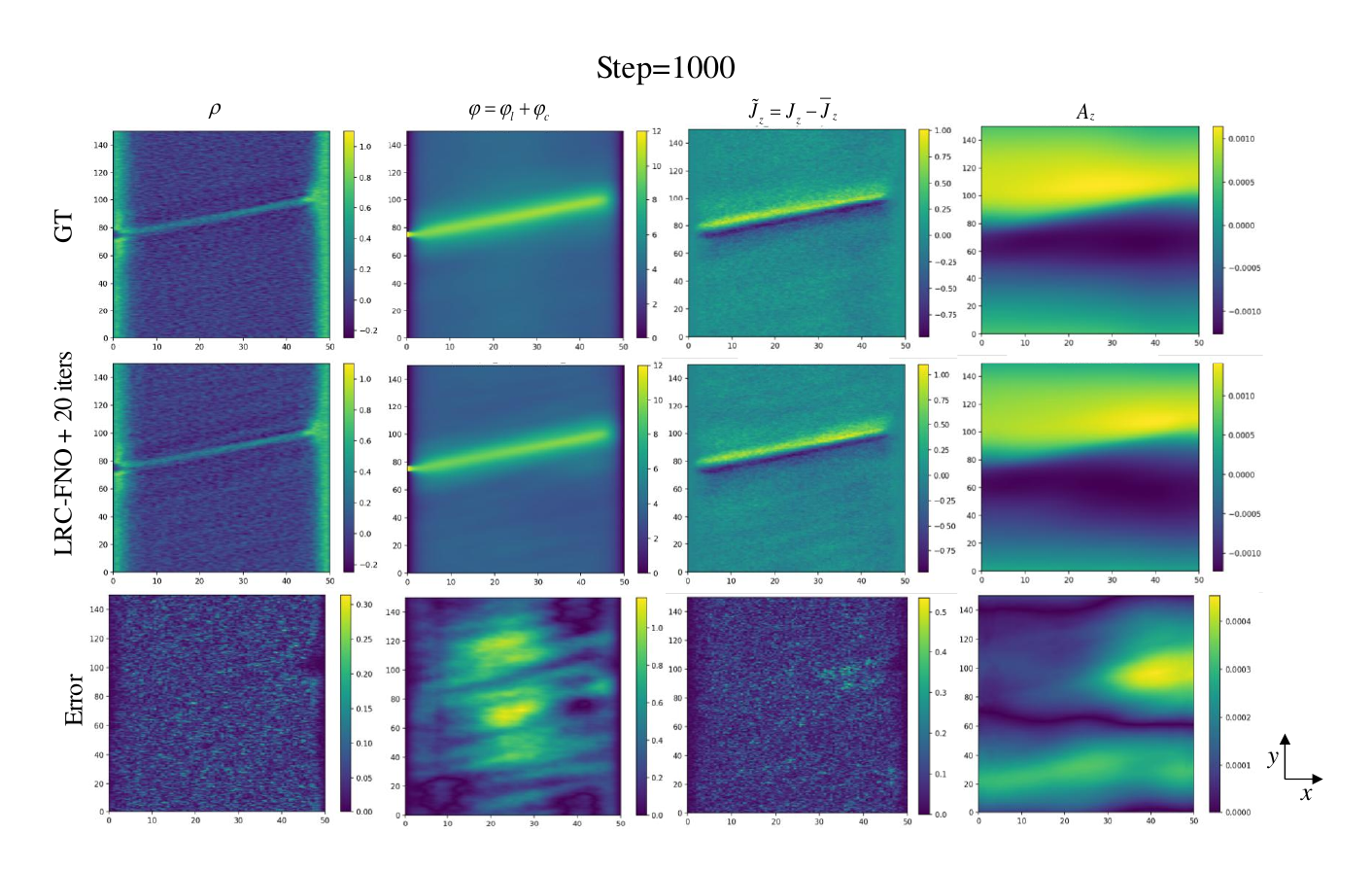}
	\caption{Multi-field distributions and errors at step = 1000 in the SOL-PIC benchmark using LRC-FNO-assisted iterative correction.}
	\label{fig:fig11}
\end{figure}

\begin{figure}[!b]
	\centering
	\includegraphics[width=15cm]{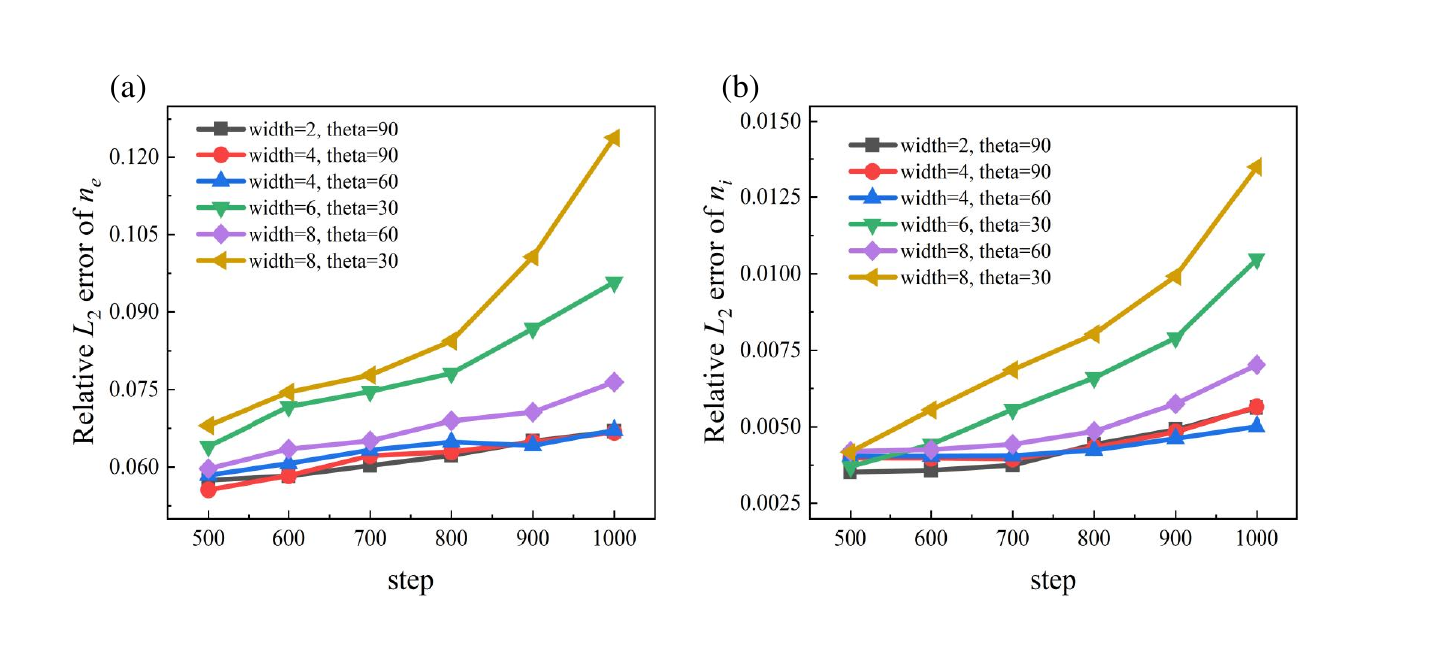}
	\caption{Evolution of relative L2 errors of electron and ion densities in extrapolated SOL-PIC simulations using LRC-FNO-assisted iterative correction.}
	\label{fig:fig12}
\end{figure}

\paragraph{}
Nevertheless, the overall charge density, ion density, and current density distributions obtained with LRC-FNO remain close to the reference solution, suggesting that the model can still capture the main spatial structures and global evolution trends of the SOL system. Therefore, although directly using LRC-FNO as the PIC field solver in this complex-boundary case may still lead to local error accumulation, its computational efficiency and ability to preserve the overall distribution indicate that it can serve as a useful tool for rapid assessment, parameter screening, and approximate prediction in SOL-PIC simulations.

\paragraph{}
Based on the above results, the LRC-FNO prediction is further used as the initial guess for the iterative field solver, followed by a fixed number of 20 iterative corrections at each time step before being used in the closed-loop SOL PIC evolution. As shown in Figure 10 with the LRC-FNO initialization, only a small number of iterations is required to obtain charge density and current density distributions that are highly consistent with the reference solution. The charge density error distribution shows no evident localized error structure near the sheath, and the y direction stripe-like errors observed in the direct LRC-FNO-driven case are effectively suppressed. This indicates that LRC-FNO can provide a high-quality initial field for the conventional iterative solver, enabling the recovery of the self-consistent field distribution under complex boundary conditions with reduced iteration cost, while preserving the particle-transport characteristics along the oblique magnetic field.

\paragraph{}
The same strategy is further extrapolated to step = 1000, as shown in Figure 11. At this stage, the overall density distributions remain close to the reference solution, although more pronounced vertical stripe-like structures appear in the potential error. This suggests that, as the extrapolation time increases, local field errors still gradually accumulate and affect the subsequent particle-field coupled evolution.

\paragraph{}
Figure~12 presents the evolution of the relative $L_2$ errors of the electron density $n_e$ and ion density $n_i$ obtained using the ``LRC-FNO initialization + 20 iterative corrections'' strategy during the extrapolated time steps. 
The 20 correction iterations are performed using the same finite-difference iterative solver as the reference field solver, with the LRC-FNO prediction used only as the initial field. 
At $\mathrm{step} = 500$, all tested cases maintain low error levels, indicating that the method can accurately reproduce the PIC physical results within the training-time range. 
As the simulation proceeds further into the extrapolation regime, the errors gradually increase and begin to grow more rapidly after $\mathrm{step} = 800$, suggesting that the iterative solver assisted by the surrogate initialization progressively deviates from the reference evolution during long-time closed-loop integration. 
Nevertheless, the method still maintains sufficient accuracy beyond the training-time range, demonstrating good robustness and extrapolation capability.

\begin{table}[htbp]
	\centering
	\caption{Computational performance comparison of different field solvers in SOL-PIC simulations.}
	\label{tab:sol_pic_runtime}
	\begin{tabular}{lcccc}
		\toprule
		& \begin{tabular}[c]{@{}c@{}}Average field solution\\time per step\end{tabular}
		& \begin{tabular}[c]{@{}c@{}}Field solver\\speedup\end{tabular}
		& PIC total runtime
		& Overall speedup \\
		\midrule
		Baseline                    & $499.65\,\mathrm{ms}$ & $1.00\times$  & $802.23\,\mathrm{s}$ & $1.00\times$ \\
		LRC-FNO                     & $14.99\,\mathrm{ms}$  & $33.33\times$ & $329.26\,\mathrm{s}$ & $2.44\times$ \\
		LRC-FNO + 20 corrections    & $28.17\,\mathrm{ms}$  & $17.74\times$ & $339.95\,\mathrm{s}$ & $2.36\times$ \\
		\bottomrule
	\end{tabular}
\end{table}

\paragraph{}
Table~2 summarizes the timing comparison between the baseline field solver and the LRC-FNO-assisted field solvers in the SOL-PIC simulations. 
The results show that, when LRC-FNO is directly used as the field solver, the average field-solution time is reduced from $499.65\,\mathrm{ms}$ to $14.99\,\mathrm{ms}$, corresponding to a field-solver speedup of $33.33\times$ and an overall PIC speedup of $2.44\times$. 
When the LRC-FNO prediction is used as the initial guess for the iterative solver followed by 20 fixed correction iterations, the average field-solution time is $28.17\,\mathrm{ms}$, still achieving a field-solver speedup of $17.74\times$ and an overall computational speedup of $2.36\times$.

\paragraph{}
These results demonstrate that LRC-FNO can substantially reduce the field-solving cost in complex-boundary SOL-PIC simulations, thereby improving the efficiency of parametric scans and multi-case comparative studies. Together with the preceding field and particle-distribution results, LRC-FNO can be used not only for fast approximate assessment, but also as a high-quality initial guess for conventional iterative solvers, reducing field-solving cost while preserving physical consistency. Therefore, the proposed method provides a useful computational tool for long-time prediction, large-scale parametric studies, and digital-twin-oriented PIC workflows.

\section{Conclusions}
\label{sec:sec5}
\paragraph{}
In this work, a Latent Residual-Closure Fourier Neural Operator (LRC-FNO) was developed for robust surrogate multi-field solving in particle-in-cell (PIC) simulations. Instead of treating field prediction as a purely data-driven regression task, LRC-FNO formulates PIC field solving as a two-level residual-closure problem involving source field compression and source-to-field operator learning. By combining latent-space closure with field-space residual correction, the proposed framework reduces the cost of repeated field solves while improving the physical consistency of closed-loop PIC integration.
\paragraph{}
The method was validated on three representative benchmarks with increasing complexity: 1D linear Landau damping, 2D two-stream instability, and a 2D scrape-off layer PIC model. In the LLD and TSI cases, LRC-FNO showed better physical consistency than the baseline neural operators by preserving potential-mode evolution, charge-density structures, and particle-field energy exchange. These results demonstrate that the two-level closure mechanism can reduce non-physical truncation errors introduced by source field compression and coarse operator learning, thereby improving the reliability of extrapolated closed-loop PIC simulations.
\paragraph{}
In the SOL case, LRC-FNO was further tested in a more practical multi-field setting involving electrostatic and magnetoquasistatic field solves. Directly embedding the network output into PIC evolution provided a fast approximate solution, while using the LRC-FNO prediction as the initial guess for the iterative solver with 20 correction iterations produced results closely consistent with the numerical benchmark. The corrected solver maintained robust density and current structures over an extrapolated time range close to twice the training horizon. In terms of efficiency, direct LRC-FNO and LRC-FNO-initialized iterative correction achieved field-solver speedups of 33.33× and 17.74×, respectively, while both strategies provide an overall PIC acceleration close to 2.5×.

\paragraph{}
Overall, LRC-FNO provides a scalable and physically consistent neural field-solving strategy for accelerating PIC simulations. It can be used either as a direct surrogate solver for fast approximate assessment or as a high-quality initial guess for conventional iterative solvers. Future work will focus on coupling LRC-FNO with implicit PIC methods, extending the framework to particle deposition and collision modules, and further developing full-process acceleration strategies for realistic electromagnetic PIC simulations.

.

\section*{Acknowledgments}
\label{sec:acknowledgments}
\paragraph{}
This work was supported in part by the National Natural Science Foundation of China (92470102, 52577152). 

\section*{Data availability}
\label{sec:data availability}
\paragraph{}
The data that support the findings of this study are available from the corresponding author upon reasonable request.

\section*{Software Implementation}
\label{sec:SOFTWARE IMPLEMENTATION}
\paragraph{}
The code used in this study is part of a larger, ongoing project and will be made publicly available upon the completion of the full project. In the meantime, readers who are interested in accessing the implementation or reproducing the results are welcome to contact the authors directly.

\bibliographystyle{unsrt}


\end{document}